\documentclass[preprint,
 amsmath,amssymb,
]{revtex4-2}
\usepackage{graphicx}
\usepackage{subcaption}
\usepackage{dcolumn}
\usepackage{bm}
\usepackage{amsmath}
\usepackage{siunitx}
\usepackage{booktabs}
\usepackage{pifont}
\usepackage{standalone}
\def\V{\mathcal{V}}

\def\P{\mathcal{P}}
\def\Nddiff{{N_{d}^{\mathrm{uniq}}}}

\usepackage{tikz}
\usetikzlibrary{shapes.geometric, arrows.meta, positioning, calc}
\tikzstyle{process} = [rectangle, rounded corners, minimum width=4.8cm, minimum height=1.2cm, text centered, draw=black, fill=blue!10]
\tikzstyle{small} = [rectangle, rounded corners, minimum width=2.2cm, minimum height=1.2cm, text centered, draw=black, fill=blue!10]
\tikzstyle{arrow} = [thick,->,>=stealth]

\begin{document}

\preprint{}

\author{Zachary T. Jerzyk}
\email{jerzyk@wisc.edu}
 \affiliation{Department of Physics, University of Wisconsin -- Madison, Madison, WI 53706, USA}

\author{David R. Smith}
\affiliation{Department of Nuclear Engineering and Engineering Physics, University of Wisconsin -- Madison, Madison, WI 53706, USA}
\email{david.smith@wisc.edu}

\author{Matthew Otten}
\email{mjotten@wisc.edu}
 \affiliation{Department of Physics, University of Wisconsin -- Madison, Madison, WI 53706, USA}
 \affiliation{Department of Chemistry, University of Wisconsin -- Madison, Madison, WI 53706, USA}

\title[Atom SHCI]
  {First ionization potentials of Cr, Mo, and W calculated with SHCI}

\date{\today}

\begin{abstract}

The design and performance of future fusion power plants will depend on accurate atomic data for plasma-facing material and plasma impurity species. A leading candidate for the plasma-facing material is tungsten due to its high melting point, however, the energy levels and wavefunctions of high-Z atoms with many electrons (e.g. 30 or more), including tungsten, are difficult to calculate with high accuracy. Gaps and large uncertainties in atomic data for tungsten introduce design and performance uncertainties for a fusion power plant. Specifically, improved atomic data for ionization potential, excited state energies, and collisional excitation rates are needed for the low charge states of atomic tungsten. We aim to address these shortcomings by using the semistochastic heat-bath configuration interaction (SHCI) method, which nearly exactly calculates the energies that can be determined at higher cost with the full configuration interaction. Adding well-motivated approximations to SHCI, including orbital optimization and effective core potentials, we demonstrate good agreement between our calculated first ionization potentials and the best available experimental values for chromium, molybdenum, and tungsten. The efficiency and accuracy achieved in calculating these ionization potentials demonstrates that our SHCI workflow can yield improved electron structure data for ions with many electrons, suggesting that the method could also be useful for collisional processes, such as state-selective charge exchange reactions and electron impact ionization.
\end{abstract}

\maketitle

\section{Introduction}


\begin{figure}[t!]
    \centering
    \includegraphics[width=\linewidth]{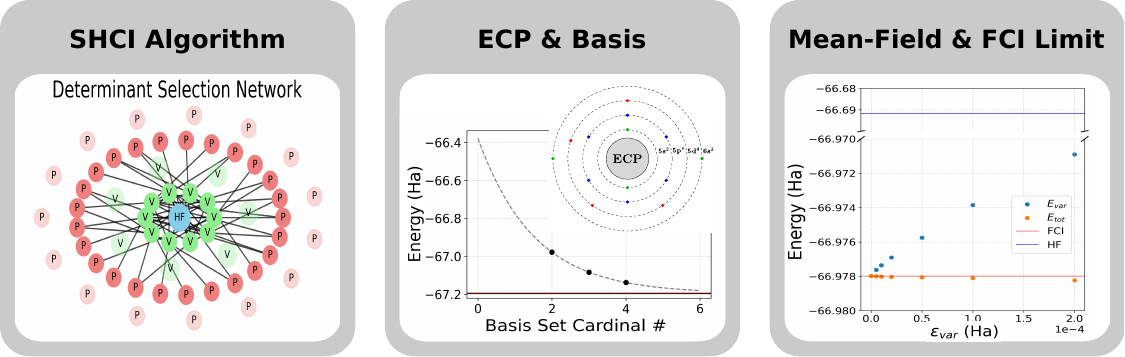}
    \caption{Summary of key steps in workflow. (Left) A visual network of determinants handled by SHCI, starting from an initial Hartree-Fock determinant and connecting to only the determinants that meet a selection criteria from a variational, then a perturbative subspace. (Middle) Visual of tungsten neutral tungsten atom with Effective Core Potential substituting core electrons, and plot of its energy converging to the limit of an ``infinitely large'' basis set. (Right) Convergence of variational and perturbatively corrected (total) energies toward the FCI Limit from the HF starting point during SHCI, over increasingly strict thresholds.}
    \label{fig1}
\end{figure}

The design and performance of magnetic fusion reactors~\cite{Shimada2007,Fasoli2023,Bigot2019,Bigot2022,Windsor2019} require accurate atomic data for plasma-facing material (PFM) and plasma impurity species.
The choice of PFM is important because high heat flux and neutron irradiation can compromise PFM, impurities from PFM can radiate energy in the core plasma, and recycling at the vessel wall generates a cold neutral population at the plasma edge.
A leading candidate for the PFM is tungsten~\cite{Khripunov2015,Beiersdorfer2015,Muller2015}, not only for its high melting point of 3695 K, but also for its relatively high resistance to sputtering.
Partially ionized plasma impurities from wall material and other impurity sources cool the plasma through radiative losses, so accurate models for impurity sources, charge states, energy levels, and collisional processes are critical for fusion plasma performance.
Current methods to calculate fusion atomic data such as the R-matrix method~\cite{Burke2011,Badnell1996,Smyth2018}, Cowan code~\cite{Quinet2010,Quinet2011} and other mean-field implementations~\cite{Badnell1996} have produced accurate eigenstates and energy levels for low-Z atoms such as carbon and neon, but atomic data for higher Z atoms, such as tungsten, are challenging due to many-electron quantum correlations~\cite{Pütterich2019}, leading to large uncertainties in spectral line identification, plasma composition, and inferred plasma density and temperature.~\cite{Quinet2010,Quinet2011}.

To address gaps in atomic data, we present a method to calculate accurate wavefunctions and the ground state energy and apply it to neutral and singly ionized chromium, molybdenum, and tungsten atoms, which are all transition metals with the same number of valence electrons. In particular, we used semistochastic heat-bath configuration interaction (SHCI)~\cite{Sharma2017}, a computationally efficient quantum chemistry approximation of full configuration interaction (FCI)~\cite{Ross1952} which is untenable for calculating energies of systems with many electrons and orbitals. Figure~\ref{fig1} (left) depicts a network of selected (connected) and unselected (isolated) determinants from both variational and perturbative subspaces, where the complete network of connections is the group of determinants represented in the final SHCI calculation. Further, Figure~\ref{fig1} (right) shows a series of SHCI calculations of increasing threshold strictness, converging from the mean-field value toward the FCI limit. Using SHCI with effective core potentials and basis set extrapolation, represented in Figure~\ref{fig1} (middle), we reduce systematic errors enough to achieve experimental accuracy and motivate the extension of these calculations to excited state energies, which are needed to calculate collisional transition rates~\cite{Kramida2006}. 

\section{Methods}

\subsection{Solving a 2$^{nd}$ quantized Hamiltonian with a given basis set \& pseudopotential}
In general, FCI and SCI methods determine the ground state of a molecular Hamiltonian with a given number of electrons, defined as
\begin{align}
    \hat{H} = \sum_{p,q} h_q^p \hat{a}_p^\dagger \hat{a}_q 
    + \frac{1}{4} \sum_{p,q,r,s} h_{qs}^{pr} \hat{a}_p^\dagger \hat{a}_r^\dagger \hat{a}_s \hat{a}_q,
\end{align}
where $\hat{H}$ is given in the second quantization formalism, which represents many-electron systems using creation $(a^\dagger)$and annihilation $(a)$ operators acting on a Fock space~\cite{Szabo1996}. In this formalism, $h_q^p$ is the one-electron Hamiltonian matrix element, and $h_{qs}^{pr}$ is the antisymmetrized two-electron matrix element, where there is a summation for repeated indices and creation and annihilation operators for electrons in the corresponding spin orbitals. The advantage of the second-quantized Hamiltonian is that it simplifies the problem by describing the occupation of states, rather than the first-quantization method of describing the states of individual particles, which introduces redundant states from identical particles.


Included in the overall Hamiltonian is a chosen representation of the electronic wavefunction, known as the basis set. The single and two-electron matrix elements $h^p_q$ and $h^{pr}_{qs}$ have an integral representation,~\cite{Helgaker2000,Szabo1996} which includes a set of functions with which to represent the probability that an electron can be found in a given location, in a particular orbital. The choice of basis set includes the number and type of orbitals which electrons may occupy and the corresponding wavefunction representations, which typically vary in the rate at which the probability distribution falls off with distance from the nucleus. Hence, a well-motivated choice of basis set will need to weigh the accuracy and number of the integrals, which contribute toward the dimensionality of the Hamiltonian, against the computational efficiency of solving a higher dimensional problem.

One way to address the overall computational efficiency and allow room for use of more complex basis sets is to substitute the ``core electrons'', that is, the electrons which occupy the innermost orbitals and are rarely ever excited to different states, with a pseudopotential that accounts for the overall Coulomb, correlation, and relativistic effects of these inactive electrons~\cite{Schwerdtfeger2011}. These kinds of  pseudopotential are known as ``Effective Core Potentials'' (ECPs)~\cite{Trail2015}, and they are also represented in the Hamiltonian matrix elements such that the single-electron element, for example, is given by
\begin{align}
    h^p_q=\langle\phi_p|-\frac{1}{2}\nabla^2+V_{ECP}|\phi_q\rangle.
\end{align}
With the inclusion of $V_{ECP}$ terms, fewer electrons are represented in the Hamiltonian matrix, and the problem is of lower dimensionality.

With a chosen basis set and ECP, the FCI energy of a system can be solved in exponential time, but even with the ECP approximation and simpler basis sets, the number of electrons and orbitals involved in the energy calculations for heavier atoms such as tungsten necessitates very large computation time, and motivate the use of an approximate algorithm, such as Selected Configuration Interaction (SCI) methods. SCI methods approximate FCI by selecting only certain important determinants with the largest coefficients from the full list of determinants included in an FCI calculation. SHCI is a specific SCI algorithm used in this work, which has been shown to achieve FCI-quality energies on systems of $10^{38}$ determinants~\cite{Li2018}.

\subsection{The SHCI Algorithm}

The SHCI algorithm~\cite{Sharma2017, Li2018} solves the many-body Schr\"{o}dinger equation in a two step process. The first is a variational step that is characteristic of SCI methods, represented in other SCI methods such as the CIPSI~\cite{Evangelisti1983} and ASCI~\cite{Tubman2016} methods, which make a determinant selection based off a set of starting determinants. Unless the system has strong correlation or near-degeneracies, this set usually includes only the Hartree-Fock determinant, which is an optimized-for Slater determinant assumed to be the best approximation for the overall wavefunction when limited to a single determinant. It is determined variationally with mean-field approximation methods. The second step is an added perturbative step that applies semistochastic perturbation theory consistent with the Heat-Bath Configuration Interaction~\cite{Holmes2016} (HCI) method.

\subsubsection{Variational Step}

Starting from an initial determinant (i.e., the Hartree-Fock determinant), SHCI generates a variational wavefunction $\Psi_V$, and iterates upon it such that the wavefunction may be represented at each step as a linear combination of determinants in a space $\V$, which represents the set of variational determinants associated with the chosen Hamiltonian,

\begin{align}
\Psi_V = \sum_{D_i \in \V} c_{i} \left|D_{i}\right\rangle .
\end{align}

At each iteration, new determinants ${D_a}$ are added to the space $\V$, where $D_a$ are determinants from the space $\P$, which represents a set of determinants that are not present in $\V$, but have a connecting Hamiltonian matrix element, $H_{ai}$, between the states $D_a$ and $D_i$. Specifically, these added determinants $D_a$ satisfy the condition that 
\begin{align}
    \exists\; D_i \in \V , \mathrm{\ such\ that\ } \left|H_{a i} c_{i}\right| \ge \epsilon_1,
\end{align}
where $\epsilon_{1}$ is a parameter chosen to specify how significant the candidate determinants need be for addition to $\Psi_V$. A choice of $\epsilon_{1} = 0$ corresponds to minimal strictness so that every possible determinant is added back into $\Psi_V$, making the calculation equivalent to FCI. This particular selection criteria is established by previous HCI and SHCI literature~\cite{Holmes2016, Sharma2017} as a much cheaper alternative to the perturbative expression used by previous SCI methods such as CIPSI~\cite{Huron1973}, which uses the full second-order perturbative correction to the energy, given by
\begin{align}
    \sum_{a \in \P} \frac{\left(\sum_{i \in \V} H_{ai} c_i\right)^2}{E_0 - E_a}.
\end{align}
By contrast, the SHCI representation uses only the simple numerator of the perturbative correction, which has a magnitude highly correlated with the significance of the associated determinant. 

When $\Psi_V$ is constructed after each selection iteration, the Hamiltonian matrix is compiled and diagonalized with the Davidson method~\cite{Davidson1989}, which iteratively estimates the lowest eigenvalue, $E_V$ and the corresponding eigenvector, $\Psi_V$ until the change in the value of $E_V$ with each iteration becomes less than a chosen threshold value.

\subsubsection{Perturbative Step}

Starting from the variational wavefunction $\Psi_V$ obtained after all iterations of Davidson convergence, Epstein-Nesbet perturbation theory~\cite{Epstein1926,Nesbet1955} provides the zeroth-order Hamiltonian, $H_o$ and perturbation, $V_{PT}$, such that 

\begin{align}
H_0 &= \sum_{i,j \in \V} H_{ij} |D_i\rangle\langle D_j| + \sum_{a \notin \V } H_{aa} |D_a\rangle\langle D_a|, \nonumber\\
V_{PT} &= H - H_0 . \label{eq:part}
\end{align}

The first and second order energy corrections are

\begin{align}
 \Delta E_1 &= 0,\nonumber\\
 \Delta E_{2} &= \langle\Psi_0|V|\Psi_1\rangle
 \;=\; \sum_{a \in \P} \frac{\left(\sum_{i \in \V} H_{ai} c_i\right)^2}{E_0 - E_a},
\label{eq:PTa}
\end{align}
where $E_a=H_{aa}$. However, calculating $\Delta E_2$ in this fashion requires a costly summation over all determinants in $\P$, so SHCI limits the terms evaluated in the inner sum to those which meet the criteria
\begin{align}
    \left|H_{a i} c_{i}\right| \ge \epsilon_2,
\end{align}
where $\epsilon_2$ is a less strict selection threshold than the $\epsilon_1$ threshold used in the variational step, that is, $\epsilon_2< \epsilon_1$. Since it is still expensive to calculate the pared down sum with a sufficiently small choice of threshold for $\epsilon_2$, the calculation is further divided into deterministic and stochastically sampled contributions to the total perturbative correction such that 
\begin{align}
    \Delta E_{2} \left(\epsilon_{2}\right) = \left[\Delta E_{2} ^{\mathrm{s}} \left(\epsilon_{2} \right) - \Delta E_{2} ^{\mathrm{s}} \left(\epsilon_{2} ^{\mathrm{d}}\right)\right] + \Delta E_{2} ^{\mathrm{d}} \left(\epsilon_{2} ^{\mathrm{d}}\right),
\label{eq:semistoch_PT}
\end{align}
where $\epsilon_2 ^{\mathrm{d}} \ge \epsilon_2$ is a third threshold value that captures the most important contributions via the deterministic component, and the remainder is made up in the stochastically sampled component. Here,
\begin{equation}
\Delta E_{2} ^{\mathrm{d}} \left(\epsilon_{2}^{\mathrm{d}}\right) = \sum_a \frac{\left(\sum_{D_i \in \V}^{(\epsilon_{2}^{\mathrm{d}})}  H_{a i} c_{i}\right) ^{2}}{E_{V} - H_{a a}}
\label{eq:PTb}
\end{equation}
and
\begin{align}
 \Delta E_{2} ^{\mathrm{s}}(\epsilon_2) =
 \frac{1}{N_{d} \left(N_{d} - 1\right)} \left\langle \sum_{D_a \in \P} \left[\left(\sum_{D_i \in \V} ^{{\Nddiff}, \left(\epsilon_2\right)} \frac{w_{i} c_{i} H_{a i}}{p_{i}}\right) ^{2} \right. \right. \nonumber \\
+ \left. \left. \sum_{D_i \in \V} ^{{\Nddiff}, \left(\epsilon_2\right)} \left(\frac{w_{i} \left(N_{d} - 1\right)}{p_{i}} - \frac{w_{i} ^{2}}{p_{i} ^{2}}\right) c_{i} ^{2} H_{a i} ^{2}\right] \frac{1}{E_{0} - E_{a}} \right\rangle
\label{eq:stoch_PT}
\end{align}
where $N_d$ is the number of variational determinants per sample and
$\Nddiff$ is the number different determinants in a sample.
$p_i$ and $w_i$ are the probability of selecting determinant $D_i$ and the number of copies of that determinant in a
sample, respectively.
The $N_d$ determinants are sampled from the discrete probability distribution
\begin{align}
p_{i} &= {\frac{\left|c_{i}\right|}{\sum_j^{N_\V} \left|c_{j}\right|}},
\label{sampling_prob}
\end{align}
using the Alias method~\cite{Walker1977,Kronmal1979}, which allows samples to be drawn
in ${\cal O}(1)$ time.
$\Delta E_2^{\mathrm{s}}[\epsilon_2]$ and $\Delta E_2^{\mathrm{s}}[\epsilon_2^{\mathrm{d}}]$ are calculated using the same set of samples,
and thus there is significant cancellation of stochastic error.

\subsection{Calculations}

For each chosen neutral atom or ion, the FCI integrals were generated with the appropriate ECP and a corresponding basis set, which were obtained from the publicly available Pseudopotential Library website~\cite{Kent2020}. The chosen ECPs are from a generation of ECPs designed for correlation-consistence, known as ccECPs~\cite{Annaberdiyev2018}, which are available for chromium, molybdenum and tungsten, and have been demonstrated to be widely applicable with good accuracy~\cite{Wang2022}.

In generating the integrals, the energy of the state was computed using a restricted open shell Kohn-Sham (ROKS), density functional theory (DFT) method~\cite{Okazaki1998} with the Local-density approximation (LDA) functional. 
We used the ROKS+DFT method following the previous work~\cite{Trail2015}, then opted to re-generate the integrals with a simple HF calculation to obtain a single starting determinant with a good overlap with the true ground state.


To accelerate convergence, we performed orbital optimization~\cite{Yao2021} at this step, which is a rotation in the orbital space which minimizes the calculated variational energy for a multi-reference wavefunction. We used an iterative optimization algorithm, the accelerated diagonal Newton method, which generates a new set of integrals for the orbital-optimized problem.
\begin{figure}[t!]
    \centering
    \includegraphics[width=\linewidth]{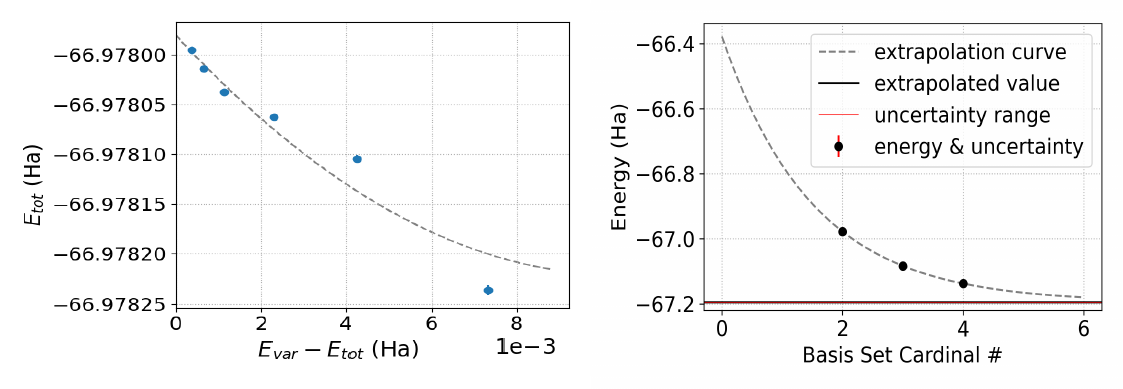}
    \caption{(left) Extrapolation of $E_{tot}$ vs. perturbative correction ($E_{var} - E_{tot}$) for neutral tungsten in aug-cc-pVDZ basis. (right) Basis set extrapolation for neutral tungsten.}
    \label{fig:extrap_summary}
\end{figure}


Since the orbital optimization is performed during an SHCI calculation, two calculations are required for each optimization to obtain the optimized integrals and then perform the final calculation. The SHCI calculations for optimization were not intended to benchmark the best possible final value, so they were configured with only a single, moderately strict threshold value for the variational step, of $\epsilon_1=2*10^{-4}$. Choosing this value saved compute time for this intermediate step, and matched the largest $\epsilon_1$ value from the list of values used in post-optimization calculations. For all calculations in both the optimizing and final steps, values of $\epsilon_2$ and $\epsilon_2^d$ were determined from the set of $\epsilon_1$ values specified in configuration such that 
\begin{align}
    \epsilon_2&=10^{-3}\epsilon_1\nonumber,\\
    \epsilon_2^d&=10^{-2}\epsilon_1\nonumber.
\end{align}

Each included $\epsilon_1$ value and the corresponding energy calculation, separated into variational and perturbative contributions, were used to extrapolate toward the $\epsilon_1=0$ limit, representing the FCI limit. This was done by fitting a second order polynomial function to the relationship between total energy and perturbative correction, in order to provide an extrapolated value of $E_{tot}$ for $E_{pt}=0$. This extrapolated value represents theoretically perfect accuracy obtained in the variational step, with no needed contribution from a perturbative correction. The extrapolated total energy was determined as the intercept, $\beta_0$, of the weighted least-squares fit associated with the polynomial function
\begin{align}
    y_i=\beta_0+\beta_1x_i+\beta_2x_i^2
\end{align}
where the weighting factor is given by 
\begin{align}
    w_i=\frac{1}{x_i^2}=\frac{1}{E_{PT}^2}\nonumber.
\end{align}
This fit and extrapolation are shown for a neutral tungsten calculation in Figure~\ref{fig:extrap_summary} (a).

Extrapolating to $\epsilon_1=0$ corresponds to our best available value for the energy of a system in a particular basis set; however, we can improve accuracy with respect to experimental values by performing an additional extrapolation with the energies calculated in different basis sets, which removes the systematic error from an incomplete basis set. We used double, triple, and quadruple-zeta basis sets published along with the corresponding ECP in the Pseudopotential Library (aug-cc-pVDZ, aug-cc-pVTZ, aug-cc-pVQZ). The extrapolation toward a value corresponding to an ``infinitely large'' basis set was performed by fitting a curve to the data plotted as basis set cardinal number versus energy, such that the final extrapolated value is given by
\begin{align}
    E_\infty=\frac{E(Dz)E(Qz)-E(Tz)^2}{E(Dz)-2E(Tz)+E(Qz)},
\end{align}
where $E(Dz)$, $E(Tz)$ and $E(Qz)$ are the energy values obtained by $\epsilon_1$ extrapolation in the double, triple and quadruple-zeta basis sets~\cite{Vasilyev2017}. The fit and extrapolation to the final value for neutral tungsten are shown in Figure~\ref{fig:extrap_summary} (b).

For certain systems, such as neutral and singly ionized tungsten, this overall workflow may lead to greater error in the ground state energy calculation if there exist nearly or exactly degenerate states that are difficult to distinguish, and can thus end up in superpositions of each other. This issue is characteristic of linear algebra calculations with degeneracies, and affects the FCI calculation as well as SCI calculations. In our case with tungsten, we attempted to distinguish the states by calculating both the ground and first excited states. This approximately doubled the SHCI computation time for both steps of our tungsten calculation, but led to greater accuracy. In general, each additional excited state calculated during a single run of SHCI should add compute time additively.


\section{Results}

The final values for the first ionization potentials, given in Table~\ref{finalvals}, were determined by taking the simple algebraic difference of the basis-set-extrapolated values for the neutral-state energy and singly-ionized-state energy of the same atom. Chromium data was obtained using the UW-Madison HEP compute cluster, and the molybdenum and tungsten data obtained using Argonne's LCRC Improv cluster, which had a higher compute power that allowed for calculations with a smaller variational threshold.

SHCI calculations for the 3d series transition metals, including chromium, are demonstrated to be accurate in molecular and atomic systems, and the method is used as a converged reference point in a direct comparison between many-body methods for various electronic wavefunctions, performed by Williams et al~\cite{Williams2020}. A 2021 analysis by Yao et al.~\cite{Yao2021_SHCI} used SHCI to calculate the ionization potential of chromium with an error of 0.006 eV, using the method by Trail and Needs~\cite{Trail2015} along with their published correlated electron pseudopotential for chromium. We used the same method with the ccECP published for chromium by Annaberdiyev et al.~\cite{Annaberdiyev2018} and achieved an error of 0.07 eV. SHCI calculations to obtain the ionization potentials of molybdenum and tungsten have not been previously published, and those presented here in Table \ref{finalvals} demonstrate much better accuracy for molybdenum than with tungsten despite both using the same generation of ccECP published by Wang et al~\cite{Wang2022}. 

Current state of the art data for tungsten in plasma applications is calculated by R-matrix methods, which are costly but effective, and have included excitation energy calculations for $W$~\cite{Smyth2018}, $W^+$~\cite{Dunleavy2022}, and $W^{2+}$~\cite{McCann2024}. Meaningful comparison to these methods would require an SHCI calculation for $W^{2+}$ as well as higher excited states for all three charge states, which is computationally feasible. 

\begin{table}[t!]
    \caption{Final ionization potentials determined by SHCI and current literature values. Note that the experimental value for tungsten reference here is one published in 1996 with a relatively small uncertainty, but certain sources more recent than 1996 reference an older experimental value of 7.98~\cite{Lide2005_Ionization} (error of 0.14 compared to our result)}
    \label{finalvals}
	\begin{center}
		\begin{tabular}{ccccc}
			\toprule
			  & Cr & Mo & W  \\
			\midrule
			SHCI (eV) & 6.70 $\pm$ 0.01 & 7.076 $\pm$ 0.002 & 8.124 $\pm$ 0.004 \\
			literature (eV) &   6.76651 $\pm$ 0.00004~\cite{Sugar1985, NIST_ASD} & 7.09243~\cite{Rayner1987}& 7.86404 $\pm$ 0.0001~\cite{Campbell-Miller1996} &  \\
            error (eV) & 0.07 & 0.016 & 0.260 \\
			\bottomrule
		\end{tabular}
	\end{center}
\end{table}

\subsection{Efficiency}
This work focused on benchmarking the best values achievable with the SHCI technique, which required significant compute resources. One of our more expensive calculations, the neutral Mo calculation in the quadruple-zeta basis set with $\epsilon_1 = 1*10^{-5}$, used approximately 7500 core hours in Argonne's LCRC Improv cluster and obtained an IP with an error of 16.97 meV

\begin{figure}[t!]
    \centering
    \begin{subfigure}{0.48\textwidth}
        \centering
        \includegraphics[width=\linewidth]{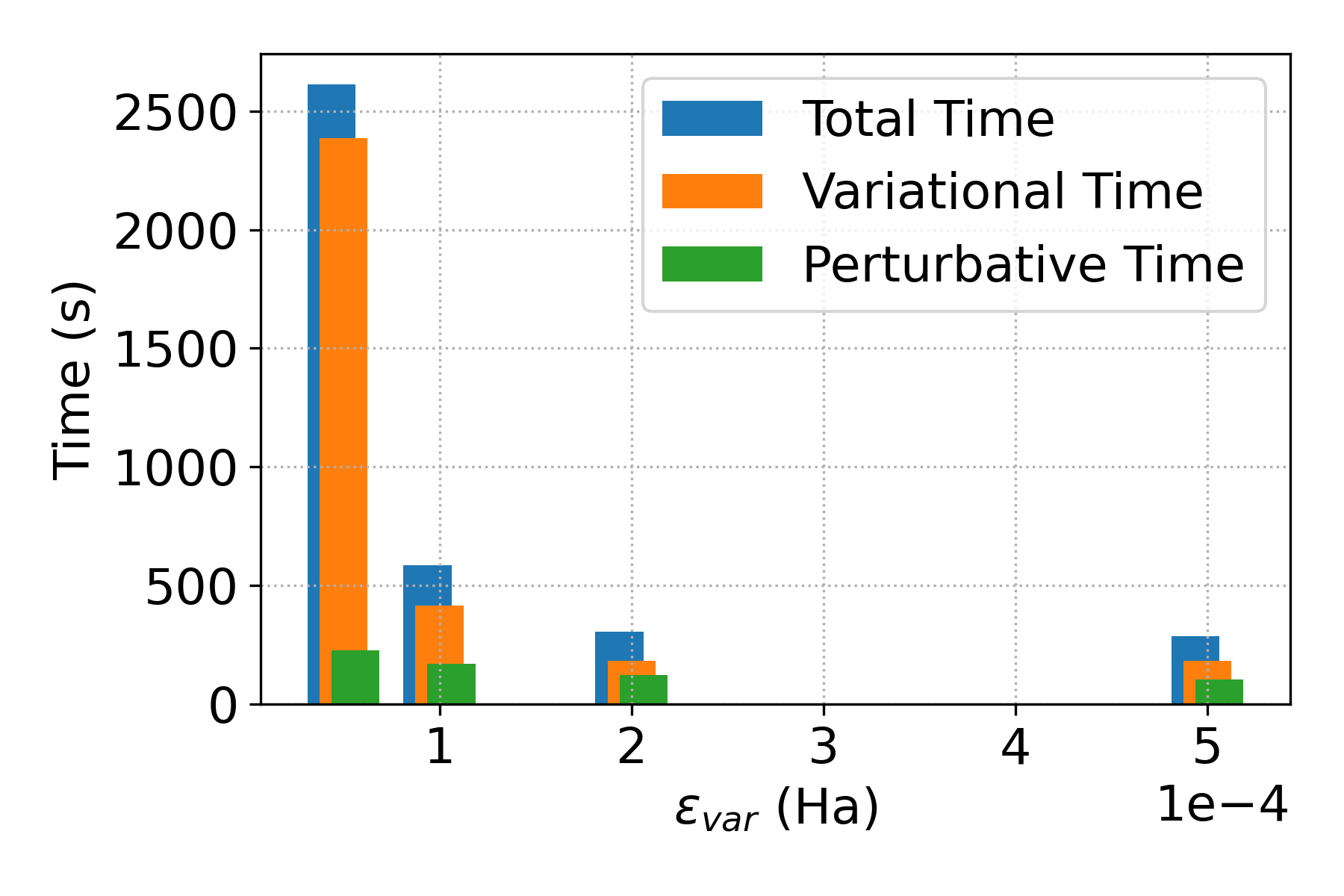}
    \end{subfigure}
    \begin{subfigure}{0.48\textwidth}
        \centering
        \includegraphics[width=\linewidth]{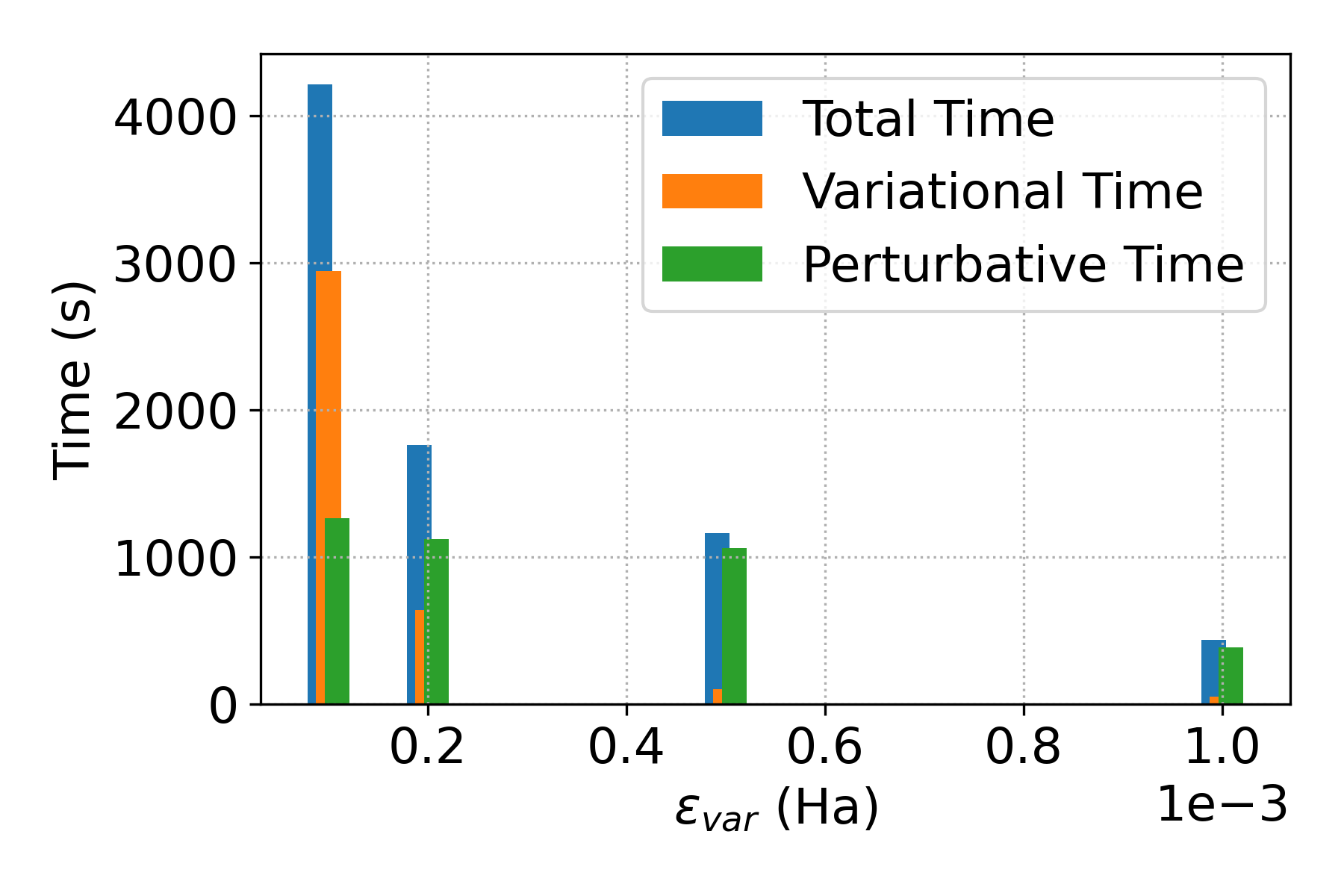}
    \end{subfigure}
    \caption{SHCI computation time per $\epsilon_1$ (eps\_var) value for an ROHF calculation of neutral chromium in aug-cc-pVQZ basis with (left) and without (right) orbital optimization}
    \label{compute_time}
\end{figure}

Despite the expenditure used in this work, we highlight that the method offers good accuracy for considerably fewer compute resources. This is demonstrated by the fact that the same neutral Mo calculation also yields an answer with an error of 11.9 meV when using a much larger threshold of $\epsilon_1 = 2*10^{-4}$. At this value, the variational step, which is the most computationally expensive step, costs approximately 10 s of compute time, compared to approximately 5200 s for $\epsilon_1 = 1*10^{-5}$. Table~\ref{mo_extrap} summarizes the results and accuracy of those tests. Additionally, Figure~\ref{compute_time} presents a visual comparison of compute time versus variational threshold for a sample calculation of neutral chromium, with and without orbital optimization.


\begin{table}[t!]
    \caption{Values calculated by SHCI for neutral molybdenum with different parameters with corresponding time, t$_{calc}$, of largest calculation step}
    \label{mo_extrap}
	\begin{center}
		\begin{tabular}{ccccc}
			\toprule
			  Method & Value (eV) & Error (eV) & t$_{calc}$ (s)\\
			\midrule
			Hartree-Fock & 6.230 & 0.862 \\
            2e-4 & 7.0805 $\pm$ 0.0009 & 0.0119 & 9.783 \\
			1e-4 & 7.0780 $\pm$ 0.0005 & 0.0144 & 29.069 \\
			5e-5 & 7.0758 $\pm$ 0.0002 & 0.0166 & 140.068\\
            2e-5 & 7.07538 $\pm$ 0.00008 & 0.01705 & 1257.317 \\
            1e-5 & 7.07546 $\pm$ 0.00007 & 0.01697 & 5191.456 \\
            extrapolated & 7.076 $\pm$ 0.002 & 0.016 & same as smallest $\epsilon_1$ used \\ 
			\bottomrule
		\end{tabular}
	\end{center}
\end{table}

\section{Discussion}

The SHCI method with ECP approximation yielded results for the first IPs of chromium and molybdenum that were within 0.1 eV, and the first IP of tungsten had an error of 0.260 eV when solving for both the ground and first excited states.

In order to push the accuracy of the tungsten value closer to the level observed for chromium and molybdenum, additional investigation into the approximations used in the tungsten model are warranted. In the near-term, there is benefit in leveraging the existing SHCI framework's efficiency to calculate higher excited states and ionized states for tungsten, since there is a need for this data to be improved for fusion purposes, especially the excited states, from which most ionizations occur for atoms in fusion conditions. Excited states will require an additive increase in compute time for each additional excited state calculated, whereas the ionized states will decrease in compute time due to fewer electrons being involved in the calculation. However, computing ionized states beyond $W^{14+}$ will require special handling, since electrons that are substituted with the ECP cannot be removed from the system. Further, it is unclear whether an ECP is as effective an approximation when the electrons it substitutes start becoming the outermost electrons. One consideration worth testing is the accuracy obtained by using an ECP designed for a smaller element, with a larger one (i.e. using the ccECP designed for Mo with W). The disadvantage of this would be the very large compute times required to make a direct accuracy comparison for low ionized states, since tungsten with a molybdenum ECP would involve 32 additional electrons in the calculation. However, if we determine that comparable accuracy is possible, then ionized states up to $W^{46+}$~\cite{Kramida2006} can be calculated with this method. For such highly ionized states, however, we may consider new approximations to account for relativistic effects, beyond those included in the ECP. If all these tests yield good results, we could even consider using the method to obtain new results for heavy atoms which do not yet have an available ccECP, by instead using the associated ECP of an element above that one on the periodic table (e.g. calculating IPs of xenon using the available ccECP for krypton).

In addition to the ionization and excitation energies, there is need of improved atomic data for tungsten's collisional cross section and radiative rates. Where this method is successful in determining an eigenstate for which the associated eigenvalue represents an accurate energy of the state, that success may be matched for other observables that are calculable from the same wavefunctions (e.g. the electric dipole moment to calculate Einstein A coefficients for radiative rate calculations). Such applications will require a recasting or reimplementation of the SHCI algorithm to be compatible for certain data, but are likely realizable in the near-term.



\section*{Acknowledgments}
We thank Adelle Wright, Margaret Fairborn, Stuart Loch, and Benjamin Faber for useful discussions.
We gratefully acknowledge the computing resources provided on Improv, a high-performance computing cluster operated by the Laboratory Computing Resource Center at Argonne National Laboratory.

\appendix

\section{Chromium}

Here we present the complete set of total ground state energy calculations for neutral and singly-ionized Cr in three basis sets (aug-cc-pVDZ, aug-cc-pVTZ, aug-cc-pVQZ), and to the level of accuracy afforded by different choices of variational threshold, $\epsilon_1$. Extrapolations in $\epsilon_1$ represent the final value determined for a given calculation.

\begin{table}[t!]
	\begin{center}
		\begin{tabular}{ccccccc}
			\toprule
             & \multicolumn{2}{c}{aug-cc-pVDZ} & \multicolumn{2}{c}{aug-cc-pVTZ} & \multicolumn{2}{c}{aug-cc-pVQZ} \\
             \midrule
			  $\epsilon_1$ (Ha) & Value (Ha) & Uncertainty (Ha) & Value (Ha) & Uncertainty (Ha) & Value (Ha) & Uncertainty (Ha)\\
			\midrule
            \num{2e-04} & -86.377846 & \num[scientific-notation=true]{0.000006} & -86.518904 & \num[scientific-notation=true]{0.000007} & -86.580236 & \num[scientific-notation=true]{0.000008}\\
            \num{1e-04} & -86.377887 & \num[scientific-notation=true]{0.000004} & -86.518949 & \num[scientific-notation=true]{0.000005} & -86.579961 & \num[scientific-notation=true]{0.000005}\\
            \num{5e-05} & -86.377950 & \num[scientific-notation=true]{0.000002} & -86.519038 & \num[scientific-notation=true]{0.000002} & -86.579820 & \num[scientific-notation=true]{0.000004}\\
            \num{3e-05} & & & & & -86.579763 & \num[scientific-notation=true]{0.000007}\\
            \num{2e-05} & -86.377983 & \num[scientific-notation=true]{0.000001} & -86.519078 & \num[scientific-notation=true]{0.000002} &  & \\
            \midrule
            extrapolated & -86.378026 & \num[scientific-notation=true]{0.000161} & -86.519138 & \num[scientific-notation=true]{0.000393} & -86.579682 & \num[scientific-notation=true]{0.000053}\\
			\bottomrule
		\end{tabular}
        \caption{Cr0 ground state total energies (calculations in the aug-cc-pVQZ basis used a slightly different set of $\epsilon_1$ values, which negligibly effects the extrapolated value)}
	\end{center}
\end{table}

\begin{table}[t!]
	\begin{center}
		\begin{tabular}{ccccccc}
			\toprule
             & \multicolumn{2}{c}{aug-cc-pVDZ} & \multicolumn{2}{c}{aug-cc-pVTZ} & \multicolumn{2}{c}{aug-cc-pVQZ} \\
             \midrule
			  $\epsilon_1$ (Ha) & Value (Ha) & Uncertainty (Ha) & Value (Ha) & Uncertainty (Ha) & Value (Ha) & Uncertainty (Ha)\\
			\midrule
            \num{2e-04} & -86.131428 & \num[scientific-notation=true]{0.000005} & -86.269512 & \num[scientific-notation=true]{0.000005} & -86.330730 & \num[scientific-notation=true]{0.000005}\\
            \num{1e-04} & -86.131424 & \num[scientific-notation=true]{0.000002} & -86.269415 & \num[scientific-notation=true]{0.000003} & -86.330478 & \num[scientific-notation=true]{0.000003}\\
            \num{5e-05} & -86.131423 & \num[scientific-notation=true]{0.000001} & -86.269364 & \num[scientific-notation=true]{0.000001} & -86.330359 & \num[scientific-notation=true]{0.000002}\\
            \num{2e-05} & -86.1314188 & \num[scientific-notation=true]{0.0000002} & -86.2693181 & \num[scientific-notation=true]{0.0000004} & -86.330274 & \num[scientific-notation=true]{0.000001}\\
            \midrule
            extrapolated & -86.131416 & \num{2e-5} & -86.269276 & \num{7e-5} & -86.330200 & \num{1e-4}\\
			\bottomrule
		\end{tabular}
        \caption{Cr1 ground state total energies}
	\end{center}
\end{table}


\section{Molybdenum}

The complete set of total ground state energy calculations for neutral and singly-ionized Mo in three basis sets (aug-cc-pVDZ, aug-cc-pVTZ, aug-cc-pVQZ), and to the level of accuracy afforded by different choices of variational threshold, $\epsilon_1$. Extrapolations in $\epsilon_1$ represent the final value determined for a given calculation.

\begin{table}[t!]
	\begin{center}
		\begin{tabular}{ccccccc}
			\toprule
             & \multicolumn{2}{c}{aug-cc-pVDZ} & \multicolumn{2}{c}{aug-cc-pVTZ} & \multicolumn{2}{c}{aug-cc-pVQZ} \\
             \midrule
			  $\epsilon_1$ (Ha) & Value (Ha) & Uncertainty (Ha) & Value (Ha) & Uncertainty (Ha) & Value (Ha) & Uncertainty (Ha)\\
			\midrule
            \num{2e-04} & -67.739912 & \num[scientific-notation=true]{0.000006} & -67.873111 & \num[scientific-notation=true]{0.000008} & -67.932393 & \num[scientific-notation=true]{0.000007}\\
            \num{1e-04} & -67.739823 & \num[scientific-notation=true]{0.000004} & -67.873109 & \num[scientific-notation=true]{0.000005} & -67.932308 & \num[scientific-notation=true]{0.000005}\\
            \num{5e-05} & -67.739814 & \num[scientific-notation=true]{0.000002} & -67.873116 & \num[scientific-notation=true]{0.000002} & -67.932261 & \num[scientific-notation=true]{0.000002}\\
            \num{2e-05} & -67.739800 & \num[scientific-notation=true]{0.000001} & -67.873082 & \num[scientific-notation=true]{0.000001} & -67.932209 & \num[scientific-notation=true]{0.000001}\\
            \num{1e-05} & -67.7397782 & \num[scientific-notation=true]{0.0000004} & -67.8730587 & \num[scientific-notation=true]{0.0000004} & -67.932185 & \num[scientific-notation=true]{0.000001}\\
            \midrule
            extrapolated & -67.739743 & \num[scientific-notation=true]{0.000025} & -67.873016 & \num[scientific-notation=true]{0.000013} & -67.932141 & \num[scientific-notation=true]{0.000017}\\
			\bottomrule
		\end{tabular}
         \caption{Mo0 ground state total energies}
	\end{center}
\end{table}

\begin{table}[t!]
	\begin{center}
		\begin{tabular}{ccccccc}
			\toprule
             & \multicolumn{2}{c}{aug-cc-pVDZ} & \multicolumn{2}{c}{aug-cc-pVTZ} & \multicolumn{2}{c}{aug-cc-pVQZ} \\
             \midrule
			  $\epsilon_1$ (Ha) & Value (Ha) & Uncertainty (Ha) & Value (Ha) & Uncertainty (Ha) & Value (Ha) & Uncertainty (Ha)\\
			\midrule
            \num{2e-04} & -67.477580 & \num[scientific-notation=true]{0.000005} & -67.611389 & \num[scientific-notation=true]{0.000006} & -67.671258 & \num[scientific-notation=true]{0.000007}\\
            \num{1e-04} & -67.477583 & \num[scientific-notation=true]{0.000002} & -67.611393 & \num[scientific-notation=true]{0.000003} & -67.671189 & \num[scientific-notation=true]{0.000004}\\
            \num{5e-05} & -67.477586 & \num[scientific-notation=true]{0.000001} & -67.611390 & \num[scientific-notation=true]{0.000002} & -67.671156 & \num[scientific-notation=true]{0.000002}\\
            \num{2e-05} & -67.4775710 & \num[scientific-notation=true]{0.0000003} & -67.6113616 & \num[scientific-notation=true]{0.0000004} & -67.671115 & \num[scientific-notation=true]{0.000001}\\
            \num{1e-05} & -67.4775630 & \num[scientific-notation=true]{0.0000002} & -67.6113439 & \num[scientific-notation=true]{0.0000003} & -67.6710915 & \num[scientific-notation=true]{0.0000003}\\
            \midrule
            extrapolated & -67.477554 & \num[scientific-notation=true]{0.000003} & -67.611319 & \num[scientific-notation=true]{0.000022} & -67.671051 & \num[scientific-notation=true]{0.000011}\\
			\bottomrule
		\end{tabular}
        \caption{Mo1 ground state total energies}
	\end{center}
\end{table}

\section{Tungsten}

The complete set of total energy calculations for neutral and singly-ionized W in three basis sets (aug-cc-pVDZ, aug-cc-pVTZ, aug-cc-pVQZ), and to the level of accuracy afforded by different choices of variational threshold, $\epsilon_1$. Extrapolations in $\epsilon_1$ represent the final value determined for a given calculation. Ground state and first exscited state energy calculations are given.

\begin{table}[t!]
    \begin{center}
        \begin{tabular}{ccccccc}
            \toprule
             & \multicolumn{2}{c}{aug-cc-pVDZ} & \multicolumn{2}{c}{aug-cc-pVTZ} & \multicolumn{2}{c}{aug-cc-pVQZ} \\
             \midrule
            $\epsilon_1$ (Ha) & Value (Ha) & Uncertainty (Ha) & Value (Ha) & Uncertainty (Ha) & Value (Ha) & Uncertainty (Ha)\\
            \midrule
            \num{2e-04} & -66.978236 & \num[scientific-notation=true]{0.000005} & -67.083893 & \num[scientific-notation=true]{0.000006} & -67.138383 & \num[scientific-notation=true]{0.000006}\\
            \num{1e-04} & -66.978104 & \num[scientific-notation=true]{0.000004} & -67.083550 & \num[scientific-notation=true]{0.000005} & -67.138191 & \num[scientific-notation=true]{0.000004}\\
            \num{5e-05} & -66.978062 & \num[scientific-notation=true]{0.000002} & -67.083451 & \num[scientific-notation=true]{0.000003} & -67.137991 & \num[scientific-notation=true]{0.000002}\\
            \num{2e-05} & -66.9780374 & \num[scientific-notation=true]{0.0000004} & -67.083396 & \num[scientific-notation=true]{0.000001} & -67.137855 & \num[scientific-notation=true]{0.000001}\\
            \num{1e-05} & -66.9780140 & \num[scientific-notation=true]{0.0000003} & -67.083348 & \num[scientific-notation=true]{0.000001} & -67.137787 & \num[scientific-notation=true]{0.000001}\\
            \midrule
            extrapolated & -66.977980 & \num[scientific-notation=true]{0.000018} & -67.083317 & \num[scientific-notation=true]{0.000119} & -67.137686 & \num[scientific-notation=true]{0.000037}\\
            \bottomrule
        \end{tabular}
        \caption{W0 ground state total energies.}
    \end{center}
\end{table}

\begin{table}[t!]
    \begin{center}
        \begin{tabular}{ccccccc}
            \toprule
             & \multicolumn{2}{c}{aug-cc-pVDZ} & \multicolumn{2}{c}{aug-cc-pVTZ} & \multicolumn{2}{c}{aug-cc-pVQZ} \\
             \midrule
            $\epsilon_1$ (Ha) & Value (Ha) & Uncertainty (Ha) & Value (Ha) & Uncertainty (Ha) & Value (Ha) & Uncertainty (Ha)\\
            \midrule
            \num{2e-04} & -66.969048 & \num[scientific-notation=true]{0.000005} & -67.074656 & \num[scientific-notation=true]{0.000006} & -67.128320 & \num[scientific-notation=true]{0.000006}\\
            \num{1e-04} & -66.969022 & \num[scientific-notation=true]{0.000004} & -67.074618 & \num[scientific-notation=true]{0.000004} & -67.128198 & \num[scientific-notation=true]{0.000005}\\
            \num{5e-05} & -66.969001 & \num[scientific-notation=true]{0.000001} & -67.074617 & \num[scientific-notation=true]{0.000002} & -67.128178 & \num[scientific-notation=true]{0.000002}\\
            \num{2e-05} & -66.9689721 & \num[scientific-notation=true]{0.0000004} & -67.074596 & \num[scientific-notation=true]{0.000001} & -67.128156 & \num[scientific-notation=true]{0.000001}\\
            \num{1e-05} & -66.9689555 & \num[scientific-notation=true]{0.0000002} & -67.0745668 & \num[scientific-notation=true]{0.0000004} & -67.1281317 & \num[scientific-notation=true]{0.0000005}\\
            \bottomrule
        \end{tabular}
        \caption{W0 first excited state total energies.}
    \end{center}
\end{table}

\begin{table}[t!]
    \begin{center}
        \begin{tabular}{ccccccc}
            \toprule
             & \multicolumn{2}{c}{aug-cc-pVDZ} & \multicolumn{2}{c}{aug-cc-pVTZ} & \multicolumn{2}{c}{aug-cc-pVQZ} \\
             \midrule
            $\epsilon_1$ (Ha) & Value (Ha) & Uncertainty (Ha) & Value (Ha) & Uncertainty (Ha) & Value (Ha) & Uncertainty (Ha)\\
            \midrule
            \num{2e-04} & -66.686070 & \num[scientific-notation=true]{0.000004} & -66.790190 & \num[scientific-notation=true]{0.000005} & -66.843136 & \num[scientific-notation=true]{0.000005}\\
            \num{1e-04} & -66.686045 & \num[scientific-notation=true]{0.000002} & -66.790139 & \num[scientific-notation=true]{0.000003} & -66.843019 & \num[scientific-notation=true]{0.000003}\\
            \num{5e-05} & -66.686028 & \num[scientific-notation=true]{0.000001} & -66.790108 & \num[scientific-notation=true]{0.000001} & -66.842937 & \num[scientific-notation=true]{0.000001}\\
            \num{2e-05} & -66.6860097 & \num[scientific-notation=true]{0.0000002} & -66.7900760 & \num[scientific-notation=true]{0.0000004} & -66.8428741 & \num[scientific-notation=true]{0.0000004}\\
            \num{1e-05} & -66.6860004 & \num[scientific-notation=true]{0.0000001} & -66.7900597 & \num[scientific-notation=true]{0.0000002} & -66.8428471 & \num[scientific-notation=true] {0.0000003}\\
            \midrule
            extrapolated & -66.685989 & \num[scientific-notation=true]{0.000001} & -66.790036 & \num[scientific-notation=true]{0.000004} & -66.842804 & \num[scientific-notation=true]{0.000003}\\
            \bottomrule
        \end{tabular}
        \caption{W1 ground state total energies.}
    \end{center}
\end{table}

\begin{table}[t!]
    \begin{center}
        \begin{tabular}{ccccccc}
            \toprule
             & \multicolumn{2}{c}{aug-cc-pVDZ} & \multicolumn{2}{c}{aug-cc-pVTZ} & \multicolumn{2}{c}{aug-cc-pVQZ} \\
             \midrule
            $\epsilon_1$ (Ha) & Value (Ha) & Uncertainty (Ha) & Value (Ha) & Uncertainty (Ha) & Value (Ha) & Uncertainty (Ha)\\
            \midrule
            \num{2e-04} & -66.686065 & \num[scientific-notation=true]{0.000004} & -66.790187 & \num[scientific-notation=true]{0.000006} & -66.843149 & \num[scientific-notation=true]{0.000006}\\
            \num{1e-04} & -66.686044 & \num[scientific-notation=true]{0.000002} & -66.790139 & \num[scientific-notation=true]{0.000003} & -66.843026 & \num[scientific-notation=true]{0.000003}\\
            \num{5e-05} & -66.686029 & \num[scientific-notation=true]{0.000001} & -66.790114 & \num[scientific-notation=true]{0.000001} & -66.842941 & \num[scientific-notation=true]{0.000001}\\
            \num{2e-05} & -66.6860102 & \num[scientific-notation=true]{0.0000002} & -66.7900777 & \num[scientific-notation=true]{0.0000004} & -66.8428758 & \num[scientific-notation=true]{0.0000004}\\
            \num{1e-05} & -66.6860003 & \num[scientific-notation=true]{0.0000001} & -66.7900598 & \num[scientific-notation=true]{0.0000002} & -66.8428481 & \num[scientific-notation=true]{0.0000003}\\
            \bottomrule
        \end{tabular}
        \caption{W1 first excited state total energies.}
    \end{center}
\end{table}
\clearpage
\bibliography{main}

\end{document}